\documentclass{aa}  
\usepackage{graphicx}
\usepackage{txfonts}
\newcommand{\lae}{\mathrel{<\kern-1.0em\lower0.9ex\hbox{$\sim$}}}
\newcommand{\gae}{\mathrel{>\kern-1.0em\lower0.9ex\hbox{$\sim$}}}
\newcommand{\ecs}{{\rm erg} \, {\rm cm}^{-2} \, {\rm s}^{-1}} 
\begin{document}
   \title{Hard X-ray Variability of AGN
 }


   \author{V. Beckmann
          \inst{1,2,3},
           S. D. Barthelmy\inst{4},
           T.J.-L. Courvoisier\inst{1,2},
	   N. Gehrels\inst{4},
           S. Soldi\inst{1,2}, 
           J. Tueller\inst{4}
           \and
           G. Wendt\inst{1}
          }

   \offprints{V. Beckmann}

   \institute{INTEGRAL Science Data Centre, Chemin d'\'Ecogia 16, 1290 Versoix, Switzerland\\
\email{Volker.Beckmann@obs.unige.ch}
     \and
   Observatoire Astronomique de l'Universit\'e de Gen\`eve, Chemin des Maillettes 51, 1290 Sauverny, Switzerland
     \and
   CSST, University of Maryland Baltimore County, 1000 Hilltop Circle,
   Baltimore, MD 21250, USA
     \and
 Astrophysics Science Division, NASA Goddard Space Flight Center, Code 661, MD 20771, USA\\ 
}

   \date{Received July 26, 2007; accepted September 12, 2007}

 
  \abstract
   {}
   {Active Galactic Nuclei are known to be variable throughout the electromagnetic spectrum. An energy domain poorly studied in this respect is the hard X-ray range above 20 keV.}
   {The first 9 months of the {Swift}/BAT all-sky survey are used to
   study the 14 -- 195 keV variability of the 44 brightest AGN. The
   sources have been selected due to their detection significance of
   $> 10 \sigma$. We tested the variability using a maximum likelihood
   estimator and by analysing the structure function.}
   {Probing different time scales, it appears that the absorbed AGN
   are more variable than the unabsorbed ones. The same applies for
   the comparison of Seyfert 2 and Seyfert 1 objects. As
   expected the blazars show stronger variability. 15\% of the
   non-blazar AGN show variability of $> 20\%$ compared to the average flux on
   time scales of 20 days, and 30\% show at least $10\%$ flux
   variation. All the non-blazar AGN which show strong variability are
   low-luminosity objects with $L_{(14-195 \, \rm keV)} < 10^{44} \rm \,
   erg \, s^{-1}$}
   {Concerning the variability pattern, there is a tendency of
   unabsorbed or type~1 
   galaxies being less variable than the absorbed or type~2 objects
   at hardest X-rays. A more solid anti-correlation is found between
   variability and luminosity, which has been previously observed in
   soft X-rays, in the UV, and in the optical domain.}

   \keywords{Galaxies: active -- Galaxies: Seyfert  -- X-rays:
               galaxies -- Surveys
               }
   \authorrunning{Beckmann et al.}

   \maketitle


\section{Introduction}

Active Galactic Nuclei (AGN) are the most prominent persistent X-ray sources in the extragalactic sky. 
X-ray observations provide a powerful tool in order to investigate the
physical conditions in the central engine of AGN. The emission in this energy band is thought to originate close
to the supermassive black hole, providing insights into the geometry
and the state of the matter. The flux and spectral variability of the sources in the hard X-rays reflect the size and physical state of the regions involved in the emission processes (see Uttley \& McHardy 2004 for a brief review).

Data of {\it EXOSAT} showed early on that the variability of AGN in the 0.1 - 10 keV range on short time scales appears to be
  red-noise in nature (\cite{McHardy}). The corresponding power spectral density functions (PSDs) can be described by a power law with index -1 to -2. The data also showed an inverse correlation between the amplitude of variability in day-long AGN X-ray light curves and the X-ray luminosity of AGN (\cite{Barr}), although Narrow Line Seyfert 1s apparently do not follow this correlation (\cite{Turner99}).
{\it RXTE}/PCA allows us to study AGN variability in the 2--20 keV range on long
time scales. This revealed that although the variability amplitudes of
AGN with different luminosities are very different on short
time-scales, they are similar on long time-scales
(\cite{Markowitz}) of about a month. {\it RXTE} data also showed that the PSDs of AGN
show a break at long time-scales according to their black hole mass (e.g. Edelson \& Nandra 1999).

Grupe et al. (2001) analysed {\it ROSAT} (0.1 -- 2.4 keV) data of AGN and showed that
the sources with steeper spectra exhibit stronger variability than
those with a hard spectrum. 
Bauer et al. (2004) showed for 136 AGN observed by {\it Chandra} within 2 Msec in the {\it Chandra} Deep Field South that $\sim 60 \%$ show signs of variability. For the brighter sources with better photon statistics even $80 - 90 \%$ showed variability in the $0.5 - 8 \rm \, keV$ energy range.

The similarity of the variability in different types of AGN suggests
that the underlying physical mechanism is the same. This does not
apply for the blazars, for which the common model is that we look 
into a highly relativistic jet. Explanations for the variability in
Seyfert galaxies include a flare/spot model in which the X-ray
emission is generated both in hot magnetic loops above an accretion
disk and in bright spots created under the loops by strong irradiation
(\cite{Czerny}), unstable accretion disks (\cite{King04}), and variable obscuration
(e.g. Risaliti et al. 2002).
A still open question is the role of long term variability at energies
above 20 keV. Observations of AGN have been performed by several
missions like {\it CGRO}/OSSE and {\it BeppoSAX}/PDS. But long-term
coverage with base lines longer than weeks is up to now only available
from the data of {\it CGRO}/BATSE, which had no imaging capabilities. 

As the Burst Alert Telescope (BAT,
Barthelmy et al. 2005) on-board {\it Swift} (\cite{Swift}) is sensitive
in the 14-195 keV energy range, it 
preferentially detects those {\it ROSAT} AGN with hard spectra, and one expects to see a
lower variability in {\it Swift}/BAT detected AGN than measurable in
average e.g. by Grupe et al. (2001) for the {\it ROSAT} data.

    In this paper we use the data of the first 9 months of the {\it
    Swift}/BAT all-sky survey to study variability of the 40 brightest
    AGN. Data analysis is described in Section~2. Two methods are applied to determine the intrinsic
    variability. Firstly, a maximum likelihood estimator
    (\cite{Almaini}) to determine the strength of variability is used. This
    approach  is similar to
    determining the 'excess variance' (\cite{Nandra97}) but allows
    for individual measurement errors. Secondly, we apply the
    structure function (\cite{StructureFunction}) in order to find
    significant variability. The results are discussed in Section~3
    and conclusions are presented in Section~4.

\section{Data analysis}

\subsection{Swift/BAT detected AGN}

The Burst Alert Telescope (BAT, Barthelmy et al. 2005) is a large
field of view ($\sim 1.5 \rm \, sr$) coded mask aperture hard X-ray
telescope. The BAT camera is a CdZnTe array of 0.5 m$^2$ with 32768
detectors, which are sensitive in the 14--195 keV energy
range. Although BAT is designed to find Gamma-ray bursts which are
then followed-up by the narrow field instruments of {\it Swift}, the almost
random distribution of detected GRBs in the sky leads to an
effective all-sky survey in the hard X-rays. The effective exposure
during the first 9 months varies over the sky from 600 to 2500 ksec. 
As shown by Markwardt et al. (2005), who also explain the survey
analysis, one expects no false detection of
sources above a significance threshold of $5.5 \sigma$. 

Within the first 9 months, 243 sources were detected with a
significance higher than $5.5 \sigma$. Among those sources, 103 are
either known AGN or have been shown to be AGN through follow-up
observations of new detections. A detailed analysis of the AGN
population seen by {\it Swift}/BAT will be given by Tueller et al. (2007). In order to study variability, we restricted our analysis to objects
which show an overall significance of $> 10 \sigma$, resulting in 44
sources. The list of objects, sorted by their name, is given in Tab.~\ref{varcheck}, together
with the average 14--195 keV count rate and the variability estimator
as described in the next Section. Among the objects are 11 Seyfert 1, 22 Seyfert 2, 5 Seyfert 1.5,
one Seyfert 1.8, one Seyfert 1.9, and 4 blazars. 
The five blazars are 3C~454.3, 4C~+71.07, 3C~273, and Markarian~421. In addition IGR~J21247+5058 is detected, which has been identified as a
radio galaxy (\cite{Masetti04}), but which might also host a
blazar core
(\cite{ClaudioIGR}). The ten brightest sources
are (according to their significance in descending order):
Cen A, NGC 4151, NGC~4388, 3C~273,
IC 4329A, NGC 2110, NGC 5506, MCG~--05--23--016,
NGC 4945, and NGC 4507. 
For all 44 objects information about intrinsic absorption is
available from soft X-ray observations. 
Among the Seyfert galaxies we see 15 objects
with $N_H > 10^{23} \rm \, cm^{-2}$, and 4 with $N_H < 10^{21} \rm \,
cm^{-2}$. 
Examples for {\it Swift}/BAT lightcurves can be found in Beckmann et
al. (2007) for the case of NGC~2992 and NGC~3081.

\subsection{Maximum Likelihood Estimator of Variability}

 Any lightcurve consisting of $N$ flux measurements $x_i$ varies due to
  measurement errors $\sigma_i$. In case the object is also intrinsically
  variable, an additional source variance $\sigma_Q$ has to be considered. The
  challenge of any analysis of light curves of variable sources is to
  disentangle them in order to estimate the intrinsic variability. A
  common approach is to use the 'excess variance' (\cite{Nandra97}, \cite{Vaughan03})
  as such an estimator. The sample variance is given by
\begin{equation}
S^2 = \frac{1}{N-1} \, \sum_{i = 1}^N (x_i - \overline{x})^2
\end{equation}
 and the excess variance is given by
\begin{equation}
\sigma_{XS}^2 = S^2 - \overline{\sigma_i^2}
\label{excessvariance}
\end{equation}
with  $\overline{\sigma_i^2}$ being the average variance
of the measurements. Almaini et al. (2000) point out that the excess
variance represents the best variability estimator only for identical
measurement errors ($\sigma_i =$ constant) and otherwise a numerical
approach should be used. 
Such an approach to estimate the strength of variability has been described by Almaini et al. and has lately been used e.g. for
analysing {\it XMM-Newton} data of AGN in the Lockman Hole (Mateos et
al. 2007).
Assuming Gaussian statistics, for a light curve with a mean
$\overline{x}$, measured errors $\sigma_i$ and an intrinsic
$\sigma_Q$, the probability density for
obtaining $N$ data values $x_i$ is given by 
\begin{equation}
p(x_i|\sigma_i,\sigma_Q) = \prod_i^N \frac{\exp (-0.5 (x_i -
  \overline{x})^2 / (\sigma_i^2 + \sigma_Q^2))}{(2 \pi)^{1/2} (\sigma_i^2 +
  \sigma_Q^2)^{1/2}}
\end{equation}

This is simply a product of $N$ Gaussian functions representing the
probability distribution for each bin. 

We may turn this around using Bayes' theorem to obtain the probability
distribution for $\sigma_Q$ given our measurements:
\begin{equation}
p(\sigma_Q|x_i,\sigma_i) = p(x_i|\sigma_i,\sigma_Q)
\frac{p(\sigma_Q)}{p(x_i)} \propto L(\sigma_Q|x_i,\sigma_i)
\end{equation}
where $L(\sigma_Q|x_i,\sigma_i)$ is the likelihood function for the
parameter $\sigma_Q$ given the data. 
This general form for the likelihood function can be calculated if one
assumes a Bayesian prior distribution for $\sigma_Q$ and $x_i$. In the
simplest case of a uniform prior one obtains
\begin{equation}
L(\sigma_Q|x_i,\sigma_i) \propto p(x_i|\sigma_i,\sigma_Q) = \prod_i^N \frac{\exp (-0.5 (x_i -
  \overline{x})^2 / (\sigma_i^2 + \sigma_Q^2))}{(2 \pi)^{1/2} (\sigma_i^2 +
  \sigma_Q^2)^{1/2}}
\end{equation}
The parameter of interest is the value of $\sigma_Q$, which gives an
estimate for the intrinsic variation we have to add in order to
obtain the given distribution of measurements. 

By differentiating, the maximum-likelihood estimate for
$\sigma_Q$ can be shown to satisfy the following, which (for a uniform
prior) is mathematically identical to a least-$\chi^2$ solution:
\begin{equation}
\sum_{i = 1}^N \frac{(x_i - \overline{x})^2 - (\sigma_i^2 +
  \sigma_Q^2)}{(\sigma_i^2 + \sigma_Q^2)^2} = 0
\end{equation}

 In the case of identical measurement errors ($\sigma_i =$ constant)
  this reduces to the excess variance described in
  Eq.~\ref{excessvariance} and in this case $\sigma_Q = \sigma_{XS}$.
We applied this method to the lightcurves with different time binning (1 day,
7 days, 20 days, 40 days). 
$\sigma_Q$ is the intrinsic variability
in each time bin, and it is larger for shorter time binning. As
expected, the
statistical error $\sigma_i$ is also
larger for shorter time bins. But the ratio between intrinsic variability
and statistical error $\sigma_Q / \sigma_i$ is smaller for shorter
time bins. 
In order to learn something about the strength of variability, we used
$S_V = 100 \% \cdot \sigma_Q / \overline{x}$, where $\overline{x}$ is the average
count rate of the source. As a control object we use the Crab. This
constant source shows an
intrinsic variability of $S_V = 2.6 \%$ (1 day binning) down to $S_V =
1.1 \%$ (40 day binning). This value might be assumed to be the
systematic error in the {\it Swift}/BAT data. 
In addition, we used lightcurves extracted at random positions in the sky. Here
the $S_V$ value does not give a meaningful result (as the average flux
is close to zero). 
But the fact that $\sigma_Q > 0$ for a random position indicates that a $\sigma_Q$ 
as large as the one for a random position cannot be attributed to intrinsic variability, but
might instead be caused by instrumental effects or due to the image
deconvolution process.
The uncorrected results are reported in Table~\ref{varcheck}. The
variability estimator $S_V$ is given in percentage $[\%]$. Note that
the variability estimator  $\sigma_Q$ usually decreases with the
length of the time bins, as does the statistical error of the
measurements. The average fluxes are listed in detector counts per
second. As the Crab shows a count rate of $0.045 \rm \, s^{-1}$, a
count rate of $10^{-4} \rm \, s^{-1}$ corresponds to a flux of
$f_{14-195 \rm \, keV} \simeq 6 \times 10^{-11} \ecs$ for a Crab-like
($\Gamma = 2.08$) spectrum.

\begin{table*}
\caption[]{Results on variability estimator following Almaini et
  al. (2000). $\overline{x}$: {\it Swift}/BAT (14 -- 195 keV) average count
  rate; $\sigma_Q$: intrinsic variability
for 1 day binned lightcurve; $S_V$: variability estimator $S_V = 100 \% \cdot \sigma_Q / \overline{x}$}
\label{varcheck}
\begin{tabular}{lccrrrr}
source name & $\overline{x}$ & $\sigma_Q$ (1 day) & $S_V$ & $S_V$ & $S_V$ & $S_V$\\
       & [$10^{-4}$ cps] & [$10^{-4}$ cps] & (1 day) & (7 day) & (20 day) & (40
       day)\\
\hline
3C 111 & $2.24 \pm 0.08$ & 1.17 & 52.2 & 12.4 & 14.7 & 9.6\\ 
3C 120 & $2.34 \pm 0.08$& 0.89 & 37.9 & 36.2 & 12.8 & 12.1\\
3C~273 & $5.05 \pm 0.06$ & 1.51 & 31.1 & 25.2 & 23.9 & 22.1\\
3C 382 & $1.59 \pm 0.07$ & 1.09 & 68.7 & 33.5 & 15.7 & -- \\
3C 390.3 & $1.76 \pm 0.07$ & 0.92& 52.4 & 36.0 & 18.9 & 7.4\\ 
3C 454.3 & $3.44 \pm 0.07$ & 2.28 & 66.8 & 57.2 & 52.3 & 38.2\\
4C +71.07 & $1.22 \pm 0.07$ & 0.82 & 67.5 & 48.4 & 29.6 & 24.3\\
Cen A  & $13.57 \pm 0.07$ & 2.26 & 16.7 & 12.6 & 12.1 & 7.4\\
Cyg A & $2.12 \pm 0.06$ & 0.96 & 45.3 & 23.3 & 18.9 & 10.6 \\
ESO103--035 & $2.03 \pm 0.09$ & 1.38 & 68.6 & 42.4 & 18.1 & 3.8\\
ESO297--018 & $0.97 \pm 0.07$ & 1.08 & 112.2 & 64.1 & 39.3 & 27.9\\
ESO506--027 & $2.31 \pm 0.07$ & 1.46 & 63.7 & 35.6 & 27.7 & 22.0\\
EXO 055620--3820.2 & $1.04 \pm 0.06$ & 1.02 & 101.2 & 63.6 & 16.2 & 21.6\\
GRS 1734--292 & $2.33 \pm 0.11$ & 1.29 & 54.7 & 37.8 & 14.9 & 13.0\\
IC4329A & $5.62 \pm 0.07$ & 1.03& 18.3 & 13.1 & 7.4 & 6.8\\
IGR J21247+5058& $2.64 \pm 0.06$& 1.06 & 40.9 & 27.4 & 24.7 & 20.0\\ 
MCG+08--11--011 & $2.01 \pm 0.09$ & 1.20 & 59.1 & 37.6 & 26.0 & --\\
MCG--05--23--016 & $3.95 \pm 0.08$ & 1.20 & 30.4 & 21.8 & 15.1 & 8.2\\
MR 2251-178 & $2.19 \pm 0.10$ & 1.49 & 68.1 & 27.7 & 13.6 & 11.8\\
Mrk 3 & $1.92 \pm 0.07$ & 0.83 & 43.1 & 25.4 & 19.5 & 14.2\\
Mrk 348 & $1.79 \pm 0.07$ & 1.11 & 62.9 & 30.8 & 32.7 & 22.9\\
Mrk 421& $1.33 \pm 0.06$ & 3.14 & 235.3 & 217.1 & 181.6 & 178.8\\
NGC 1142 & $1.59 \pm 0.07$ & 0.79 & 50.4 & 34.4 & 18.3 & 18.5\\
NGC 1275 & $2.05 \pm 0.08$ & 1.24 & 60.1 & 29.4 & 24.6 & 17.7\\
NGC 1365 & $1.33 \pm 0.06$ & 0.88 & 66.0 & 36.4 & 31.5 & 24.8\\
NGC 2110 & $4.64 \pm 0.07$ & 1.67 & 36.0 & 31.7 & 33.3 & 32.3\\
NGC 2992 & $1.19 \pm 0.08$ & 1.34 & 111.3 & 73.8 & 76.0 & 51.9\\
NGC 3081 & $1.67 \pm 0.08$ & 1.25 & 73.1 & 46.5 & 44.4 & 23.1\\
NGC 3227 & $2.44 \pm 0.06$ & 0.78 & 32.2 & 28.3 & 20.0 & 21.8\\
NGC 3281 & $1.54 \pm 0.08$ & 0.95 & 64.4 & 36.6 & 26.8 & 23.7\\
NGC 3516 & $1.98 \pm 0.06$ & 0.93 & 47.2 & 18.8 & 10.6 & 7.7\\
NGC 3783 & $3.27 \pm 0.07$ & 1.05 & 32.1 & 16.0 & 9.0 & 7.2\\
NGC 4051 & $0.81 \pm 0.06$ & 0.81 & 99.4 & 48.3 & 16.0 & 29.0\\
NGC 4151 & $7.13 \pm 0.06$ & 2.87 & 40.3 & 35.8 & 33.3 & 30.3 \\
NGC 4388 & $4.73 \pm 0.06$ & 1.56 & 33.0 & 24.4 & 18.9 & 18.4\\
NGC 4507 & $3.51 \pm 0.07$ & 1.10 & 31.2 & 13.0 & 13.3 & 12.3\\
NGC 4593 & $1.60 \pm 0.07$ & 1.04 & 64.1 & 33.8 & 22.2 & 15.4\\
NGC 4945 & $3.66 \pm 0.07$ & 1.65 & 45.2 & 33.9 & 30.5 & 22.7\\
NGC 5506 & $4.28 \pm 0.07$ & 1.03 & 24.1 & 12.5 & 9.2 & 6.6\\
NGC 5728 & $1.67 \pm 0.08$ & 1.16 & 68.4 & 20.5 & 17.0 & 14.7\\
NGC 7172 & $2.65 \pm 0.10$ & 1.44 & 53.8 & 48.3 & 25.2 & 21.9\\
NGC 7582 & $1.19 \pm 0.08$ & 0.99 & 80.8 & 61.9 & 50.4 & 38.7\\
QSO B0241+622 & $1.41 \pm 0.07$ & 0.95 & 66.8 & 39.8 & 28.4 & 16.2\\
XSS J05054--2348& $1.07 \pm 0.06$ & 0.99 & 93.8 & 55.8 & 34.1 & 34.2\\[0.2cm]

Crab   & $453.8 \pm 0.10$ & 11.6 & 2.56 & 1.72 & 1.27 & 1.07\\
\end{tabular}
\end{table*}

The sources extracted at random positions show a $\sigma_Q = 3.6
\times 10^{-5}$ on the 20 day time scale ($\sigma_Q = 9.2 \times
10^{-5}$, $5.4 \times
10^{-5}$, $2.2 \times
10^{-5}$, for 1, 7 and 40 day binning, respectively). Thus, in
  order to get corrected for systematic errors, we subtracted $3.6
\times 10^{-5}$ from the  $\sigma_Q$ of each source in the 20 day
measurement and determined the $S_{Vc}$ based on this value: $S_{Vc} =
100 \% \, (\sigma_Q - 3.6
\times 10^{-5}) / \overline{x}$.
The errors on the variability estimator have been determined by
Monte-Carlo simulations. The flux and error distribution of each
source have been used. Under the assumption that the source fluxes and
errors are following a Gaussian distribution, for each source 1000
lightcurves have been simulated. Each of these lightcurves contains
the same number of data points as the original lightcurve. The data
have then been fitted by the same procedure and the error has been determined
based on the $1 \sigma$ standard deviation of the $\sigma_Q$ values
derived. 
The results
are shown in Table~\ref{varcheck2}. The sources have been sorted by
source type and then in descending variability. 

It has to be taken into account that the source type
  in Table~\ref{varcheck2} is based on optical observations only. The radio properties
  are not taken into account. 3C 111, 3C~120, 3C
  382, and 3C 390.3 are
  not standard Seyfert~1 galaxies but broad-line radio galaxies, and
  Cen~A and Cyg A are narrow-line radio galaxies. In the case of
  IGR~J21247+5058 the nature of the optical galaxy is not clear yet. In
  all of these cases, the prominent jet of the radio galaxy might contribute to the hard X-ray
  emission. In
  addition the optical classification is often but not always
  correlated with the absorption measured in soft X-rays: Most, but
  not all, Seyfert~2 galaxies show strong absorption ($N_H > 10^{22}
  \rm \, cm^{-2}$), whereas most, but not all, Seyfert~1 galaxies
  exhibit small hydrogen column densities ($N_H < 10^{22}
  \rm \, cm^{-2}$), as noted e.g. by Cappi et al. (2006).

\begin{figure}
\centering
\includegraphics[width=7cm]{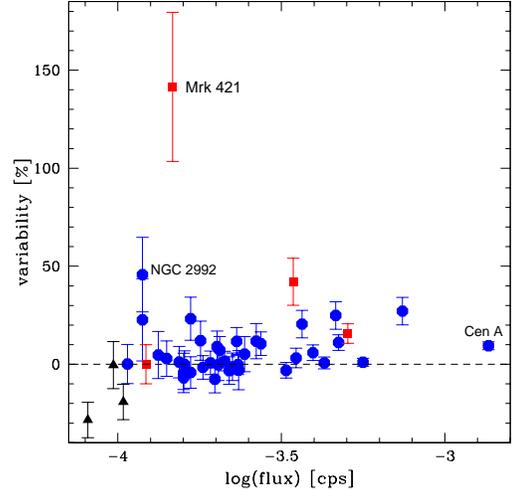}
\caption{Variability estimator $S_{Vc}$ as a function of {\it
    Swift}/BAT 14--195 keV count rate. A count rate of $10^{-4} \rm \,
  s^{-1}$ corresponds to a flux of about 
$f_{14-195 \rm \, keV} \simeq 6 \times 10^{-11} \ecs$. Blazars have been marked with
    squares and the three objects with the lowest count rates ($<
    1.05 \times 10^{-4} \rm \, s^{-1}$) are marked with triangles.}
              \label{fig:fluxvar}
\end{figure}

\begin{figure}
\centering
\includegraphics[width=9cm]{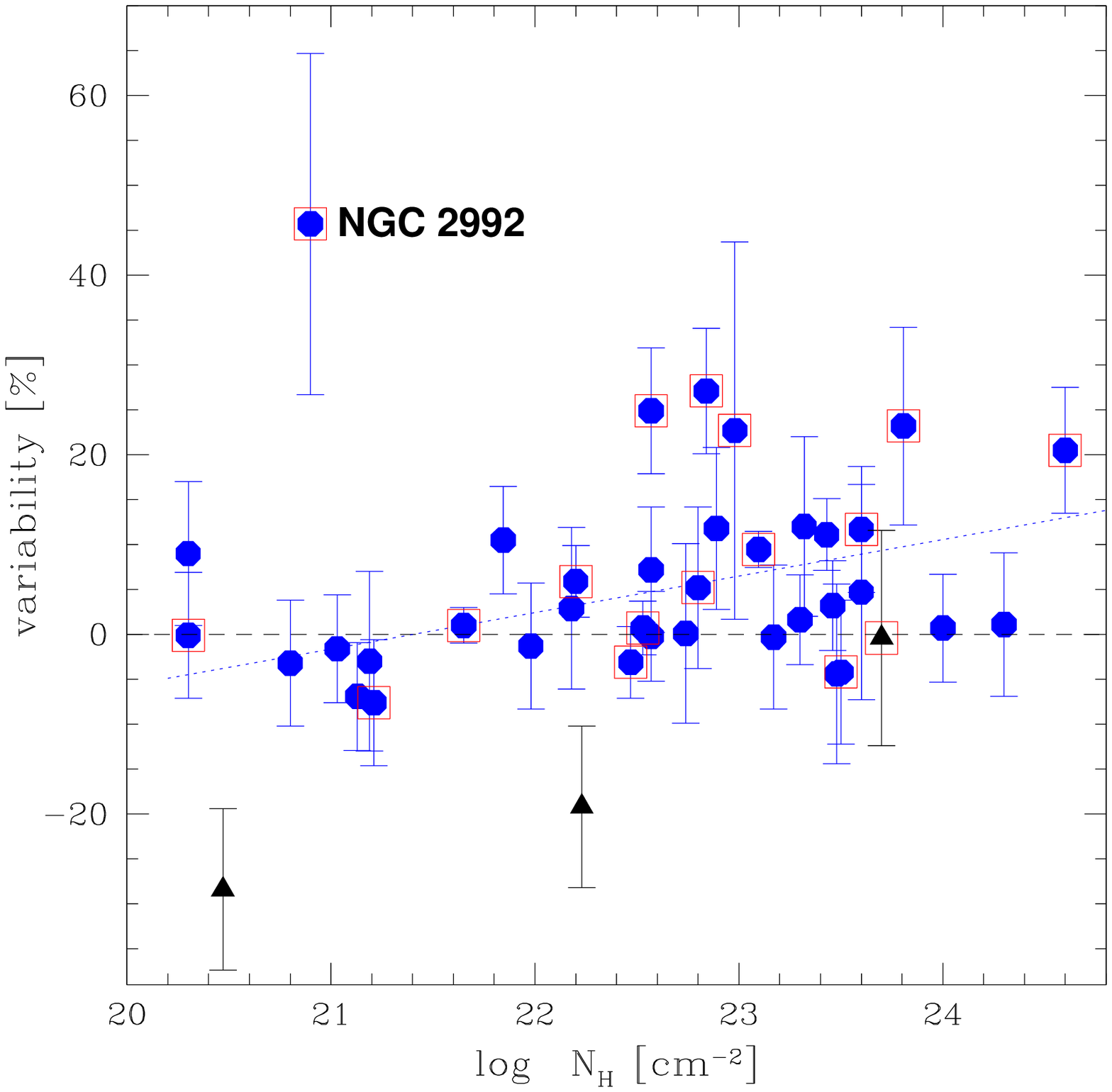}
\caption{Variability estimator $S_{Vc}$ as a function of intrinsic
  absorption $N_H$. Blazars have been excluded and the three objects with the lowest count rates ($<
    1.05 \times 10^{-4} \rm \, s^{-1}$) are marked with
    triangles. Objects enclosed by a square show a rising part of the
    structure function with a correlation probability of $rr_{SF} \le
    0.01$ (see Section~\ref{SF} for details). The
    object with the highest $S_{Vc}$ is NGC~2992.  The dotted line indicates the linear
  regression to the data points, excluding NGC~2992 and the three
  low-flux objects.}
              \label{fig:nhvar}
\end{figure}

\begin{table*}
\caption[]{Results on variability estimator - corrected values and
  structure function. {\it type}: optical classification; $\overline{x}$: {\it Swift}/BAT (14 -- 195 keV)
  average count rate; $N_H$: 
intrinsic absorption; $L_X$: luminosity (14 -- 195 keV) assuming a
Crab-like spectrum ($\Gamma \simeq 2.1$); $\sigma_Q$: intrinsic variability
for 20 day binned lightcurve;
$S_{Vc}$: corrected intrinsic variability;  $rr_{SF}$: probability for
non-correlation from structure function analysis}
\label{varcheck2}
\begin{tabular}{lcclcccc}
source name & type & $\overline{x}$ & $\log N_H$ & $\log L_X$ & $\sigma_Q$ (20 day) &
       $S_{Vc}$ & $\log rr_{SF}$ \\
       &      & [$10^{-4}$ cps] & $\rm [cm^{-2}]$ & [erg s$^{-1}$] & [$10^{-4}$ cps] &
       (20 day) [\%] & \\
\hline
Mrk 421& blazar & 1.33 & 19.0$^{1}$ & 44.21 &2.44 & $142 \pm 38$ & $-10.8$ \\
3C 454.3 & blazar & 3.44 & 20.8$^{2}$ & 47.65 & 1.81 & $42 \pm 12$ & $-3.9$ \\
3C~273 & blazar & 5.05 & 20.5$^{3}$ & 46.26 & 1.16 & $15 \pm 5$ & $-2.8$ \\
4C +71.07 & blazar & 1.22 & 21.0$^{3}$ & 48.10 & 0.36 & $0 \pm 10$ & $0.0$\\
IGR J21247+5058& rad. gal. & 2.60 & 21.8$^{4}$ & 44.15 & 0.65 & $11 \pm 6$ & $-1.1$\\
QSO B0241+622 & Sy1 & 1.41 & 22.2$^{5}$ & 44.58 & 0.40 & $3 \pm 9$  & $-0.3$\\
IC4329A     & Sy1 & 5.62 & 21.7$^{3}$ & 44.29 & 0.42 & $1 \pm 2$ & $-3.5$\\
NGC 4593    & Sy1 & 1.60 & 20.3$^{5}$ & 43.24  & 0.36 & $0 \pm 7$ & $-3.1$\\
GRS 1734--292 & Sy1 & 2.33 & 22.6$^{6}$ & 44.16 & 0.36 & $0 \pm 5$ &  $-0.7$\\
3C 111      & Sy1 & 2.24 & 22.0$^{3}$ & 44.86 & 0.33 & $-1 \pm 7$ & $-1.2$\\
3C 390.3    & Sy1 & 1.76 & 21.0$^{3}$ & 44.88 & 0.33 & $-2 \pm 6$ & $-0.2$\\
3C 120      & Sy1 & 2.34 & 21.2$^{3}$ & 44.54 & 0.29 & $-3 \pm 10$ & $-1.8$\\
NGC 3783    & Sy1 & 3.27 & 22.5$^{5}$ & 43.62 & 0.26 & $-3 \pm 4$ & $-3.6$\\
MR 2251-178 & Sy1 & 2.19 & 20.8$^{3}$ & 45.09 & 0.29 & $-3 \pm 7$ & $-0.1$\\
3C 382      & Sy1 & 1.59 & 21.1$^{3}$ & 44.86 & 0.25 & $-7 \pm 6$ & $-0.8$\\
EXO 055620--3820.2 & Sy1 & 1.04 & 22.2$^{3}$ & 44.21 & 0.16 & $-19 \pm 9$ & $-0.1$\\
NGC 4151    & Sy1.5 & 7.13 & 22.8$^{6}$ & 43.02 & 2.38 & $27 \pm 7$ & $-7.6$\\
MCG+08--11--011 & Sy1.5 & 2.01 & 20.3$^{5}$ & 44.06 & 0.54 & $9 \pm 8$ & $-0.1$\\
NGC 3227    & Sy1.5 & 2.44 & 22.8$^{7}$ & 42.69 & 0.49 & $5 \pm 9$ & $-4.2$\\
NGC 3516    & Sy1.5 & 1.98 & 21.2$^{3}$ & 43.32 & 0.21 & $-8 \pm 7$ & $-4.3$ \\
NGC 4051    & Sy1.5 & 0.81 & 20.5$^{3}$ & 41.77 & 0.13 & $-28 \pm 9$ & $-1.3$\\
NGC 1365 & Sy1.8 & 1.33 & 23.6$^{5}$ & 42.72 & 0.42 & $5 \pm 12$ & $-0.8$\\
NGC 5506 & Sy1.9 & 4.28 & 22.5$^{5}$ & 43.34 & 0.39 & $-4 \pm 3$ & $-6.1$\\
MCG--05--23--016 & Sy1.9 & 3.94 & 22.2$^{8}$ &  43.58 & 0.60 & $6 \pm 4$ & $-2.2$\\
NGC 2992 & Sy2 & 1.19 & 20.9$^{9}$ & 42.98 & 0.91 & $45 \pm 19$ & $-6.6$\\
NGC 2110 & Sy2 & 4.64 & 22.6$^{3}$ & 43.58 & 1.52 & $25 \pm 7$ & $-2.9$\\
NGC 3081 & Sy2 & 1.67 & 23.8$^{10}$ & 43.15 & 0.75 & $23 \pm 11$ & $-6.2$\\
NGC 7582 & Sy2 & 1.19 & 23.0$^{3}$ & 42.64 & 0.63 & $23 \pm 21$ & $-2.2$\\
NGC 4945 & Sy2 & 3.66 & 24.6$^{5}$ & 42.24 & 1.11 & $21 \pm 7$ & $-2.8$\\
Mrk 348 & Sy2 & 1.79 & 23.3$^{11}$ & 43.74 & 0.58 & $12 \pm 10$ & $-1.9$\\
NGC 7172 & Sy2 & 2.65 & 23.9$^{3}$ & 43.43 & 0.68 & $12 \pm 9$ & $-0.3$\\
ESO506--027 & Sy2 & 2.31 & 23.8$^{12}$  & 44.29 & 0.63 & $12 \pm 7$ & $-3.9$\\
NGC 4388 & Sy2 & 4.73 & 23.4$^{13}$ & 43.65 & 0.89 & $11 \pm 4$ & $-1.5$\\
Cen A  & Sy2& 13.57 & 23.1$^{14}$ & 42.78 & 1.65 & $10 \pm 2$ & $-2.8$ \\ 
NGC 1275 & Sy2 & 2.05 & 22.6$^{14}$ &  43.93 & 0.51 & $7 \pm 7$ & $-1.1$\\
NGC 4507 & Sy2 & 3.51 & 23.5$^{5}$ & 43.82 & 0.48 & $3 \pm 5$ & $-0.9$\\
Cyg A & Sy2 & 2.12 & 23.3$^{15}$ & 44.96 & 0.40 & $2 \pm 5$ & $-1.5$\\
NGC 3281 & Sy2 & 1.54 & 24.3$^{16}$ & 43.37 & 0.38 & $1 \pm 8$ & $-0.3$\\
Mrk 3 & Sy2 & 1.92 & 24.0$^{5}$ & 43.67 & 0.38 & $1 \pm 6$ & $-0.1$\\
XSS J05054--2348& Sy2 & 1.07 & 22.7$^{12}$ & 44.25 & 0.36 & $0 \pm 10$ & $-1.1$\\
ESO103--035 & Sy2 & 2.03 & 23.2$^{3}$ & 43.68 & 0.36 & $0 \pm 8$ & $-1.7$\\
ESO297--018 & Sy2 & 0.97 & 23.7$^{12}$ & 43.92 & 0.36 & $0 \pm 12$ & $-2.4$\\
NGC 1142 & Sy2 & 1.59 & 23.5$^{12}$ & 44.25 & 0.29 & $-4 \pm 10$  & $-0.6$\\
NGC 5728 & Sy2 & 1.67 &  23.5$^{17}$ & 43.29 & 0.29 & $-4 \pm 8$ & $-5.3$\\
\end{tabular}

References. (1) \cite{Mrk421NH}; (2) \cite{3C454NH}; (3) Tartarus
database; (4) \cite{ClaudioIGR}; (5) \cite{Lutz04}; (6)
\cite{NGC4151}; (7) \cite{NGC3227NH}; (8) \cite{Soldi05},
(9) \cite{NGC2992}; (10) \cite{NGC3081NH}; (11)
\cite{AkylasNH}; (12) from {\it Swift}/XRT analysis; (13)
\cite{NGC4388}; (14) \cite{intagn}; (15) \cite{Young02}; (16)
\cite{NGC3281}; (17) Mushotzky (private communication)
\end{table*}
 Obviously, the fainter the source is,
the more difficult it is to get a good measurement for the
variability. Thus, one might suspect that there is a correlation between source
flux and variability $S_{Vc}$. Figure~\ref{fig:fluxvar} shows the
variability estimator $S_{Vc}$ as a function of flux (14 - 195 keV in
counts per second). There is no correlation between flux
and variability, although all the sources for which no variability was
detectable are of low flux. A Spearman rank test
  (\cite{Spearman}) gives a
  correlation coefficient as low as $r_s = 0.2$, 
  rejecting the hypothesis that flux
  and variability are correlated.
The estimation of variability becomes more uncertain for objects with
very low fluxes. We therefore mark the three sources with the lowest
flux in the figures and do not consider them when studying
correlations between parameters.  
From Table~\ref{varcheck2} it is already apparent that none of the 11
type 1 galaxies shows significant variability, whereas of the 20
type 2 objects 50\% show variability with $S_{Vc} \ge 10\%$. This
effect is also apparent when comparing the variability $S_{Vc}$ with
the intrinsic absorption $N_H$ as measured in soft X-rays (e.g. by
{\it Swift}/XRT or {\it XMM-Newton}). The correlation is shown in
Figure~\ref{fig:nhvar}. Blazars have been excluded. Except for
NGC 2992, none of the objects with intrinsic absorption $N_H < 10^{22}
\rm \, cm^{-2}$ shows significant variability according to the maximum
likelihood estimator. NGC~2992 is also a special case
because it is a Seyfert~2 galaxy with comparably low intrinsic absorption and the $N_H$ varies between $0.1$
and $1.0 \times 10^{22} \rm \, cm^{-2}$ (Beckmann et al. 2007).
  Even when including NGC~2992 a
  Spearman rank test of $N_H$ versus variability gives a correlation
  coefficient of $r_s = 0.31$, which corresponds to a moderate
  probability of correlation of $95\%$.

\subsection{Structure Function}
\label{SF}

As an independent test for variability, we determined the structure
function of the objects. 
Structure functions are similar to auto- and cross-correlation
functions and have been introduced for analysis of radio lightcurves
by Simonetti et al.~(1985). Applications to other data sets have been shown,
e.g. by Hughes et al. (1992), Paltani (1999), de Vries (2005), and Favre et al. (2005).
The structure function is a useful
and simple to use tool in order to find characteristic time scales for
the variations in a source. We use the first-order structure
  function, which is defined as $D^1(\tau) = <[S(t) -
S(t+\tau)]^2>$. Here $S(t)$ is the flux at time $t$, and $\tau$ is the
time-lag, or variability time-scale. The function $D^1(\tau)$ can be
characterized in terms of its slope: $b = d \log D^1 / d \log
\tau$. For a stationary random process the structure function is
related simply to the variance $\sigma^2$ of the process and its
autocorrelation function $\rho(\tau)$ by  $D^1(\tau) = 2 \sigma^2 [1 -
\rho(\tau)]$. 
For lags longer than the
longest correlation time scale, there is an upper plateau with an amplitude
equal to twice the variance of the fluctuation ($2 \cdot (\sigma_Q^2 +
\overline{\sigma_i} ^2)$). For very short time
lags, the structure function reaches a lower plateau which is at a
level corresponding to the measurement noise ($2 \cdot
\overline{\sigma_i} ^2$). As
explained in Hughes et al. (1992), the structure function,
autocorrelation function, and power spectrum density function (PSD) $P(\nu)$ are related
measures of the distribution of power with time scale. If the PSD
follows a power law of the form  $P(\nu) \propto \nu^{-a}$, then
$D^1(\tau) \propto \tau^{a-1}$ (\cite{Bregman90}). For example, if
$P(\nu) \propto \nu^{-1}$, then $D^1(\tau) \propto \tau^0$ (flicker noise). Flicker noise exhibits both short and long time-scale
fluctuations. If $P(\nu) \propto \nu^{-2}$, then $D^1(\tau) \propto
\tau^1$ (short or random walk noise). This relation is however valid
only in the limit $\tau_{max} \rightarrow \infty$, $\tau_{min}
\rightarrow 0$. If, on the contrary, the PSD is limited to the range
$[\tau_{min},\tau_{max}]$, the relationship does not hold anymore
(\cite{Paltani99}). This is in fact the case here, as we can probe
only time scales in the range of $\tau_{min} \sim 10 \rm \, days$ to $\tau_{max} \sim 100 \rm \, days$.
  In the ideal case we can learn from the structure function of the {\it
  Swift}/BAT AGN about several physical properties: whether the
  objects show variations, what the maximum time scale of variations
  is,
  and what the type of noise is which is causing the variations. 
  We can determine the maximum time scale $\tau_{max}$ of
  variability only if a plateau is
  reached and, in our case, if $\tau_{max} < 9$ months.

  Error values on the structure function have been again determined
  by Monte-Carlo simulation. The flux and error distribution of each
source has been used. Under the assumption that the source fluxes and
errors are following a Gaussian distribution, for each source 1000
lightcurves with 1 day binning have been simulated. These lightcurves have than been used
  to extract the structure function. The scatter in each point
  $D^1(\tau)$ is then considered when fitting a straight line to the
  data applying linear regression.

\begin{figure}
\centering
\includegraphics[width=7cm,angle=90]{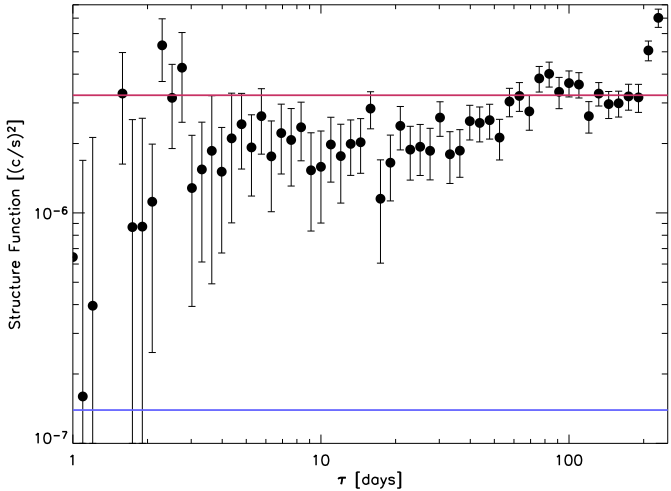}
\caption{Structure function for {\it Swift}/BAT data of the Crab (one day
  binning). The upper line indicates an amplitude
equal to twice the variance of the fluctuation, the lower line
corresponds to the
level of the average measurement noise ($2 \cdot
\overline{\sigma_i} ^2$).}
              \label{fig:Crab}
\end{figure}

\begin{figure}
\centering
\includegraphics[width=7cm,angle=90]{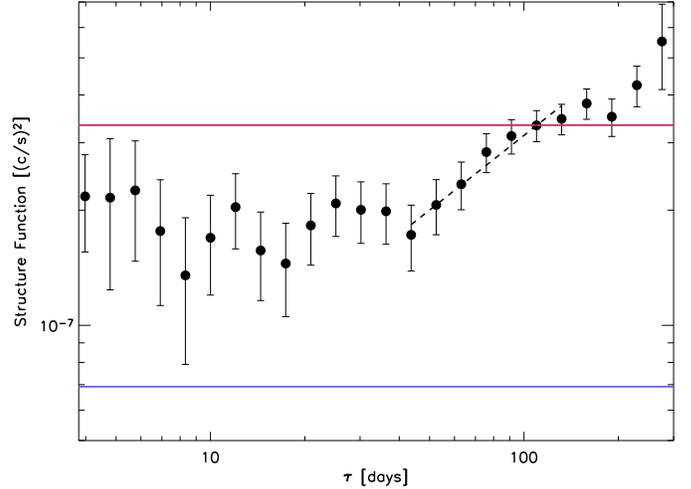}
\caption{Structure function for {\it Swift}/BAT data of the Seyfert 1.5
  galaxy NGC~4151 (one day
  binning). The upper and lower lines indicate $2 (\sigma_Q^2 +
  \overline{\sigma_i}^2)$ and $2
  \overline{\sigma_i} ^2$, respectively. The dashed line indicates the
  linear regression applied to the data, resulting in $D^1(\tau) \propto
\tau^{0.65}$.}
              \label{fig:NGC4151}
\end{figure}

\begin{figure}
\centering
\includegraphics[width=6.5cm,angle=90]{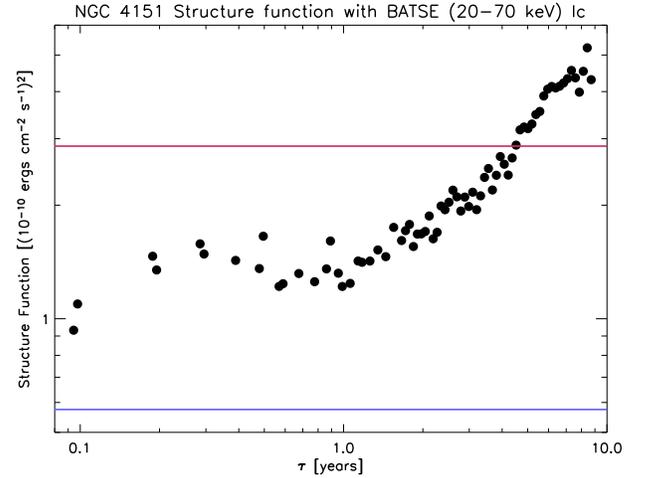}
\caption{Structure function for {\it CGRO}/BATSE data (20 - 70 keV) of
  NGC~4151. The upper and lower lines indicate $2 (\sigma_Q^2 +
  \overline{\sigma_i}^2)$ and $2
  \overline{\sigma_i} ^2$, respectively.}
              \label{fig:NGC4151BATSE}
\end{figure}

\begin{figure}
\centering
\includegraphics[width=7cm,angle=90]{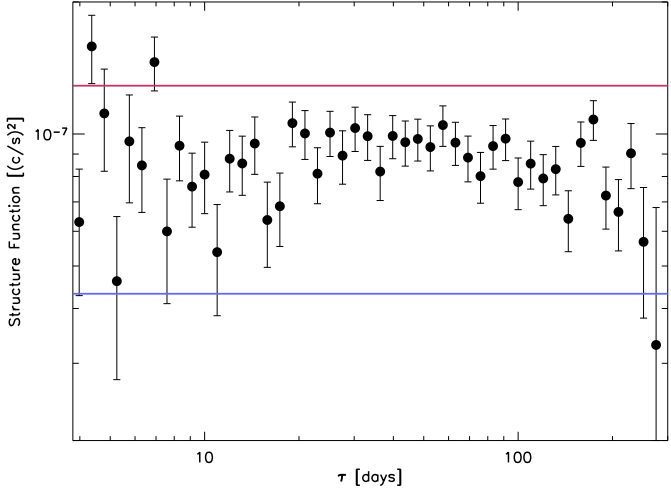}
\caption{Structure function for {\it Swift}/BAT data of a random
  position in the sky (XXX~J0044.1+5019). The upper and lower lines indicate $2 (\sigma_Q^2 +
  \overline{\sigma_i}^2)$ and $2
  \overline{\sigma_i} ^2$, respectively.}
              \label{fig:SFrandom}
\end{figure}

To test the quality of the BAT data lightcurves for determining the
structure function we show in 
Figure~\ref{fig:Crab} the one obtained for the Crab as an example for
a constant source. As expected, after the structure function
gets out of the noisy part at time scales shorter than $\sim 4$ days,
it stays more or less constant. Thus, no variability is detected in
the Crab up to time scales of the duration of the survey. 
Figure~\ref{fig:NGC4151} shows the structure function for the BAT
data of NGC~4151. This source has a lower flux than the Crab, thus the
noisy part of the structure function extends up to $\tau \simeq 20 - 
40$ days. At longer time scales, the function is rising. It is not
clear though whether it levels out after 150 days, which would mean
there is no variability on time scales longer than 150 days. But the
source is variable on timescales ranging from 3 weeks to (at least) 5
months.
For comparison we checked the {\it CGRO}/BATSE Earth-occultation
archive\footnote{http://f64.nsstc.nasa.gov/batse/occultation/} which contains light curves for 4 sources of our sample,
i.e. 3C~273, Cen~A, NGC~4151, and NGC~1275. 
Figure~\ref{fig:NGC4151BATSE} shows the $20 - 70 \rm \, keV$ structure
function of NGC~4151 based on {\it CGRO}/BATSE data. The sampling here is worse at the time scales probed by the {\it Swift}/BAT survey,
but reaches out to time scales up to $\tau \simeq 8$ years. One can see that
the turnover does not appear within the probed time scale, consistent with
the results we derived from the BAT data.
Also for the other 3 objects the results from BATSE and BAT are
consistent, showing variability for Cen~A and 3C~273 over all the
sampled time scales, and no variability for NGC~1275.
We also checked the lightcurves for random positions in the sky. One
example is shown in Figure~\ref{fig:SFrandom}. 


Based on the comparison of the structure function curves of the BAT
AGN with those of the Crab and the random positions, we 
examined the curves of all objects of the sample presented here for rising evolution in the range $\tau = 20 - 200
\rm \, days$. Individual time limits $l_{min}$ and $l_{max}$ have been
applied in
order to apply a linear regression fit to the curves, taking into
account the errors determined in the Monte-Carlo simulation. Therefore this
method inherits a subjective element which obviously limits the
usefulness of the output. On the other hand, a fixed $l_{min}$ and
$l_{max}$ does not take into account the difference in significance
between the sources. The $l_{max}$ applied is not necessarily the
maximum time-scale of variability $\tau_{max}$, especially when $\tau_{max} > 100
\rm \, days$.
The last 
column of Table~\ref{varcheck2} reports the results. $rr_{SF}$ gives
the probability for a non-correlation of $\tau$ and $D^1(\tau)$. We
consider here objects with $\log rr_{SF} \le -2$ as variable,
i.e. objects where we find a probability of $>99\%$ for
correlation. The structure function of the Crab lightcurve for example
results in  $\log rr_{SF} = -0.6$.
One can see an overall agreement with the variability estimator, although in some cases there are
 discrepancies, e.g. for NGC~3516 and NGC~5728, which have a rising structure function, but do not give
an indication of variability in the maximum likelihood approach. In
total, 16 objects show a rising structure function,
and 15 objects show a variability $S_{Vc} > 10\%$ in the maximum
likelihood approach. 
10 objects show a rising structure function and $S_{Vc} > 10\%$. 
  A Spearman rank test of the variability estimator versus the $\log
  rr_{SF}$ value gives a probability of $>98\%$ for correlation, and
  $>99.5\%$ if we ignore the objects with a negative variability
  estimator. Some caution has to be applied when comparing those two
  values: while the variability estimator measures the strength of the
  variability, the  $\log
  rr_{SF}$ indicates the probability that there is indeed significant
  variation. A bright source can have a small but very significant
  variability. The fact that the variability estimator is based on
  20-day binned lightcurves, while the structure functions are
  extracted from 1-day binned data should not affect the results
  strongly: because of the moderate sampling of the light curves, the
  structure function analysis cannot probe variability below $\sim 20
  \rm \, days$ in most cases.

Concerning a dependence of variability on intrinsic absorption, the
structure function method confirms the tendency seen in variability estimator. As shown in Fig.\ref{fig:nhvar}, $25\%$ of the
objects with $N_H < 10^{22} \rm \, cm^{-2}$ and $46\%$ of the objects
with $N_H > 10^{22} \rm \, cm^{-2}$ show a rising structure
function. Again, this should be taken as a tendency, not as a strong correlation.

\section{Discussion}

Studying the correlation between absorption and variability, 
there is a tendency that 
the stronger absorbed sources are the more variable ones
(Fig.~\ref{fig:nhvar}). If the central engine in type~1 and type~2 objects is indeed similar, this is a surprising result. First,
absorption should not play a major role in the spectrum at energies $>
15 \rm \, keV$ unless the absorption is $N_H \gg 10^{23} \rm \,
cm^{-2}$. But most of the sources studied here show only moderate
absorption with hydrogen column densities of the order of $N_H =
10^{21} - 10^{23} \rm \, cm^{-2}$. Even if absorption plays a role,
the expected effect would be reverse to the observation, i.e. one
would expect a damping effect of the absorption and the absorbed
sources should be less variable than the unabsorbed
ones. In a recent study of {\it XMM-Newton} data of AGN in the Lockman
Hole by Mateos et al. (2007) it has been shown that although the
fraction of variable sources is higher among type-1 than in type-2
AGN, the fraction of AGN with detected spectral variability were found
to be $\sim 14 \pm 8 \%$ for type-1 AGN and $34 \pm 14 \%$ for type-2
AGN. This might indicate that the differences between type~1 and
type~2 galaxies are indeed more complex than just different
viewing angles resulting in a difference
in the absorbing material along the line of sight. In this context,
alternative and modified accretion models might be considered, such as
matter accretion via clumps of matter and interaction between these
clumps (\cite{Courvoisier05}) or star collisions in a cluster of stars
orbiting around the central massive black hole (\cite{Torricelli00}).

\begin{figure}
\centering
\includegraphics[width=8.3cm]{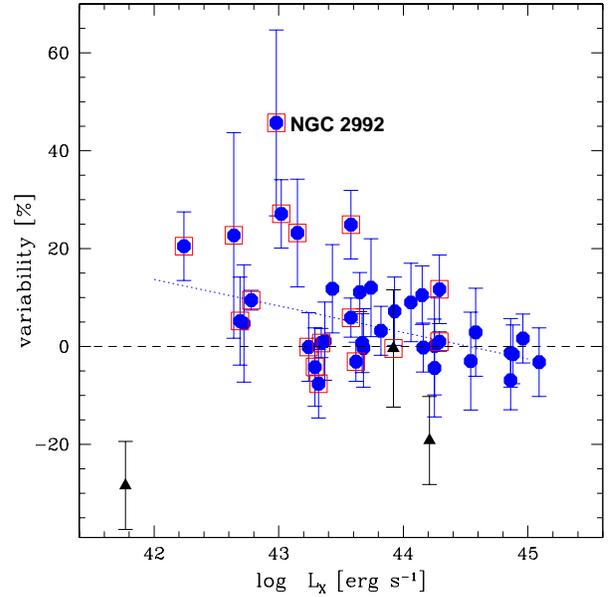}
\caption{Variability  estimator $S_{Vc}$ as a function of X-ray
  luminosity in the 14--195 keV band. The three objects with the lowest count rates ($<
    1.05 \times 10^{-4} \rm \, s^{-1}$) are marked with triangles. The blazars are located outside
  the area covered in this plot. The dotted line indicates the linear
  regression to the data points.}
              \label{fig:Lxvariability}
\end{figure}

Another explanation for the lack of variable type~1 objects in our
sample could be that the correlation between absorption
and variability is an indirect one, caused by two other correlations:
an anti-correlation of intrinsic absorption and luminosity, and the
anti-correlation of variability and luminosity. While the first
dependence in the data set presented here is very weak, there is
indeed a trend of lower variability for sources with higher
luminosity (Fig.~\ref{fig:Lxvariability}). A Spearman rank test
  of luminosity versus variability estimator results in a correlation
  coefficient of $r_s = -0.47$, which corresponds to a correlation probability of $>99\%$.
All the sources which show a $S_{Vc} >
20\%$ have luminosities of $L_{(14 - 195 \rm \, keV)} < 4 \times 10^{43} \rm \,
erg \, s^{-1}$, and all sources with $S_{Vc} >
10\%$ have  $L_{(14 - 195 \rm \, keV)} < 2\times 10^{44} \rm \,
erg \, s^{-1}$. Using the results from the structure function a
similar trend is seen: 76\% 
of the objects with $L_{(14 - 195 \rm \, keV)} < 4 \times 10^{43} \rm \,
erg \, s^{-1}$ have a significant rising part of the structure function,
whereas only 13\%
of the more luminous objects show this indication for variability.

The results based on the structure function have to be interpreted
with caution due to the relatively small number of significant data
points. Nevertheless it appears that the maximum time length
$\tau_{max}$ for variability is significantly longer than in the
optical and UV region. Collier \& Peterson (2001) studied 4 of the
objects presented here and found a $\tau_{max}$ significantly smaller
in all cases for the optical and UV. The same applies for the AGN
variability study performed by Favre et al. (2005) using UV data,
including 7 of the objects studied here. On the contrary, de Vries et
al. (2005) do not find a turn-over in optical lightcurves up to
$\tau_{max} \sim 40 \rm \, yr$.

The average gradient $\beta$ of the rising part of the structure
  functions (assuming $D^1(\tau) \propto \tau^{\beta}$) with $rr_{SF}
  < 0.01$ is $\overline{\beta} = 0.4 \pm 0.1$ and ranging for the individual
sources from $\beta = 0.2$ (3C~454.3, NGC~5506, and 3C~273) to $\beta = 1.0$
  (ESO~506-027), consistent with measurements of the power spectrum of
  11 AGN in the X-rays by {\it EXOSAT}, which resulted in $\overline{\beta} =
  0.55 \pm 0.09$ (\cite{Lawrence93}). The average value is closer to the slope
  expected from disk instability models ($\beta = 0.8 - 1.0$,
  \cite{DImodel}), rather than to the slope of the starburst model ($\beta = 1.4
  - 1.8$, \cite{SBmodel}). This result should not be overemphasized as
  de Vries et al. (2004) pointed out that the
  measurement noise does have a direct effect on the slope of the
  structure function. The larger the noise, the shallower the slope.

Compared to softer X-rays, the Seyfert galaxies appear to exhibit less
variability than e.g. at 2--10 keV. This indicates that there is an
overall tendency for an anticorrelation of variability with
energy. This has been reported for some of the objects studied here
e.g. for 3C~390.3 and 3C~120 (\cite{Gliozzi02}) which show no
significant variation here, and also for NGC~3227 (Uttley \& McHardy
2005). In the latter article the case of NGC~5506 is also described in which this trend is reversed in the soft X-rays. This object does not show significant variability applying the maximum likelihood estimator, but shows indeed a rising structure function with $\tau_{max} \gae 200$ days. 

The fraction of variable objects in our study is about $30\%$ among the Seyfert
type AGN according to both methods, the variability estimator and the
structure function. This is a lower fraction than detected at
softer X-rays. For example among the AGN in the {\it Chandra} Deep
Field South 60\% of the objects show variability (\cite{Bauer04}), and
{\it XMM-Newton} data of the Lockman Hole reveal a 50\% fraction
(\cite{Mateos07}). Part of the lower variability detected in the {\it
  Swift}/BAT AGN sample might be due to the lower statistics
apparent in the lightcurves when compared to the soft X-ray
data. Bauer et al. (2004) pointed out that the fraction of variable
sources is indeed a function of source brightness and rises up to $80
- 90\%$ for better photon statistics and also Mateos et al. find
$>80\%$ of the AGN variable for the best quality light curves. Within our sample we are
not able to confirm this trend, which might be due to the small size
of the sample.

An anticorrelation of X-ray variability with luminosity in AGN has
been reported before for energies $< 10 \rm \, keV$ (e.g. Barr \& Mushotzky 1986, Lawrence \&
Papadakis 1993) and has been also seen in the UV range
  (\cite{Paltani94}) and in the optical domain (\cite{deVries05}), although narrow-line Seyfert 1 galaxies apparently
show the opposite behaviour (e.g. Turner et al. 1999). As only one of
the objects (NGC 4051) discussed here is a NLSy1 galaxy, we detect a continuous effect
from soft to hard X-rays, which indeed indicates that the dominant underlying
physical process at $\sim5 \rm \, keV$ is the same as at $\sim20 \rm
\, keV$. 
In a more recent study, Papadakis (2004) reported that this correlation is in fact based on the connection between luminosity and the mass of the central black hole $M_{BH}$. This may be explained if more luminous sources are physically larger in size, so that they are actually varying more slowly. Alternatively, they may contain more independently flaring regions and so have a genuinely lower amplitude. The observed correlation might reflect the anticorrelation of variability and black hole mass. In the case of the sample presented here, such an anticorrelation is not detectable, but it has to be pointed out that estimates for $M_{BH}$ are only available for 13 objects. In addition, the range of objects in luminosity and black hole mass might be too small in order to detect such a trend. 
Uttley \& McHardy (2004) explained the anti-correlation of variability
and $M_{BH}$ by assuming that the X-rays are presumably produced in
optically thin material close to the central black hole, at similar
radii (i.e. in Schwarzschild radii, $R_S$) in different AGN. As $R_S =
2 G M_{BH} c^{-2}$, longer time scales for the variability
are expected for the more massive central engines, making the objects less variable on a monthly time scale
studied here.

\section{Conclusions}

We presented the variability analysis of the brightest AGN seen by
{\it Swift}/BAT, using two ways of analysis: a maximum-likelihood
variability estimator and the structure function. Both methods
  show that $\sim 30\%$ of the Seyfert type AGN exhibit significant
  variability on the time scale of $20 - 150 \rm \, days$.
The analysis indicates that the type~1 galaxies are less variable
than the type~2 type ones, and that unabsorbed sources are less
variable than absorbed ones. With higher significance we detect an anti-correlation of luminosity and variability. No object with
luminosity $L_X > 5 \times 10^{43} \rm \, erg \, s^{-1}$ shows strong
variability. The anti-correlation might either be caused by intrinsic differences between
the central engine in Seyfert 1 and Seyfert 2 galaxies, 
or it might be connected to
the same anti-correlation seen already at softer X-rays, in the UV and in the optical band. Further investigations on this
subject are necessary in order to clarify whether one can treat the
AGN as an upscale version of Galactic black hole systems (e.g. Vaughan,
Fabian \& Iwasawa 2005). 

The data presented here do not allow a final conclusion on this
point. The correlations are still too weak and for too many objects it is
not possible to determine the strength of the intrinsic
variability. Similar studies at softer X-rays seem to indicate
  that with increasing statistics we will be able to detect
  significant variability in a larger fraction of objects. The study presented here will be repeated as soon as
significantly more {\it Swift}/BAT data are available for analysis. As
this study was based on 9 months of data, a ten times larger data set
will be available in 2012. Eventually, the data will allow more
sophisticated analysis, such as the construction of power density spectra. In addition the same analysis can be
applied to {\it INTEGRAL} (\cite{INTEGRAL}) IBIS/ISGRI data. Although {\it INTEGRAL}
does not achieve a sky coverage as homogeneous as {\it Swift}/BAT, it
allows a more detailed analysis of some AGN in specific regions,
e.g. along the Galactic plane. 

The combination of results from both missions, {\it Swift} and {\it
  INTEGRAL}, should allow us to verify whether indeed Seyfert 2 galaxies
  are more variable at hard X-rays than the unabsorbed Seyfert 1, and
  whether this points to intrinsic differences in the two AGN types.

\begin{acknowledgements}
We thank the anonymous referee who gave valuable advice which helped
us to improve the paper.
This research has made use of the NASA/IPAC Extragalactic Database
(NED) which is operated by the Jet Propulsion Laboratory and of data
obtained from the High Energy Astrophysics Science Archive Research
Center (HEASARC), provided by NASA's Goddard Space Flight Center. This
research has also made use of the Tartarus (Version 3.1) database, created by Paul O'Neill and Kirpal Nandra at Imperial College London, and Jane Turner at NASA/GSFC. Tartarus is supported by funding from PPARC, and NASA grants NAG5-7385 and NAG5-7067.
\end{acknowledgements}

\end{document}